\documentclass[amsmath,amssymb,aps,prl,superscriptaddress,nobibnotes,twocolumn,floatfix]{revtex4-1}
\usepackage{times}
\usepackage{slashed}
\usepackage{graphicx}
\usepackage{dcolumn}
\usepackage{bm}
\usepackage{blindtext}

\begin{document}

\title{Thermal contribution to measured $g$-factors in alkali-like atoms}
%\author{T. Zalialiutdinov$^{1,2}$, Y. Kozhedub$^{1}$ and D. Solovyev$^{1,2}$}

%\affiliation{ $^1$Department of Physics, St. Petersburg State University, Petrodvorets, Oulianovskaya 1, 198504, St. Petersburg, Russia
%\\
%$^2$Petersburg Nuclear Physics Institute named by B.P. Konstantinov of National Research Centre "Kurchatov Institut", St. Petersburg, Gatchina, 188300, Russia
%}

\author{T. Zalialiutdinov}
\email[E-mail: ]{t.zalialiutdinov@spbu.ru}
\affiliation{Department of Physics, St. Petersburg State University, Petrodvorets, Oulianovskaya 1, 198504, St. Petersburg, Russia}
\affiliation{Petersburg Nuclear Physics Institute named by B.P. Konstantinov of National Research Centre "Kurchatov Institut", St. Petersburg, Gatchina, 188300, Russia}

\author{Y. Kozhedub}
\affiliation{Department of Physics, St. Petersburg State University, Petrodvorets, Oulianovskaya 1, 198504, St. Petersburg, Russia}

\author{D. Solovyev}
\affiliation{Department of Physics, St. Petersburg State University, Petrodvorets, Oulianovskaya 1, 198504, St. Petersburg, Russia}
\affiliation{Petersburg Nuclear Physics Institute named by B.P. Konstantinov of National Research Centre "Kurchatov Institut", St. Petersburg, Gatchina, 188300, Russia}

\date{\today}

\begin{abstract}
In this paper, we consider a one-loop thermal correction to the $g$-factors of the ground and low-lying excited states in neutral alkaline-like atomic systems. As an example, we carried out calculations for Rb and Cs atoms, where optical measurements of transition frequencies set the metrological level of accuracy. With the development of atomic systems as the most accurate tool for spectroscopic experiments, it has become increasingly important to improve the accuracy of measuring the electron $g$-factor, which is currently at ~$10^{-12}$. The rapid progress in this field in recent years requires analysis not only of various relativistic, QED and other effects, but also theoretical studies of phenomena stimulated by the external thermal environment. We can expect that the thermal contribution to the $g$-factor calculated in our work may be of interest in the near future.
\end{abstract}


\maketitle

\section{Introduction}

Drawing attention to many-electron atomic systems has a long history, starting with the first steps of quantum theory in the last century. Advances in theoretical calculations combined with success in experimental techniques have led to the most significant pride of such research, known as the atomic clock. Due to the highest frequency measurement accuracy \cite{AtCl-Cs,AtCl-Sr}, combined with the indispensable compactification \cite {Martin}, atomic clocks have found civil applications that allow precise geodesy and are of particular interest for fundamental physics research \cite{AtCl-2018}.

A discussion of the spectroscopy advantages on the many-electron systems and, in particular, on the alkali metals, can be found in \cite{Kozlov-RevModPhys,battesti_2018}. One of the possible physical applications corresponds to the measurement of the fine structure constant and Land\'{e}-factor, which can serve as a test of the atomic structure theory \cite{veseth_1987}. The theoretical study of the atomic $g$-factor has been carried out for many years, which regularly requires the matching of numerical calculations with experimental data, see, for example, Refs.~\cite{veseth_1987, Charlotte}. The latter for many-electron systems demonstrates the need to improve theoretical methods \cite{Werth_2014}. {\it Ab initio} calculations of the $g$-factors and corrections to them obtained within the framework of relativistic quantum electrodynamics (QED), being simpler in some aspects, are in agreement with experimental results for hydrogen-like ions \cite{haefner:00:prl,nature2014, oxygen,sturm:13:pra,nature2014} with an accuracy of $10^{-12}$ \cite{Hanneke}, and the measurements of  $g$-factors in highly charged Ar ions are consistent with the corresponding theory up to a comparable level of accuracy \cite{arapoglou:19:prl,micke:20:n}. 

In turn, studies of physical quantities determined with such extraordinary accuracy should be considered from the point of view of the possible influence of external effects. In particular, the thermal environment is of pronounced interest for precise measurements performed with many-electron heavy atoms, see, for example, Refs.~\cite{Nicholson,Saf-RevModPhys}. Heat-induced frequency shifts require precise theoretical calculations coupled with the development of experimental techniques aimed at comprehensive temperature control and its variation, see Ref.~\cite{Kozlov-RevModPhys} and references therein. It has recently been shown that, apart from determining the frequency and stabilizing the atomic clock, thermal corrections to the $g$-factors of bound electron approach the current level of accuracy, making them important for future prospects in this field \cite{gfactor2022}.

The obvious success in measuring the Land\'{e}-factors of a bound electron in light ions stimulated the development of optical spectroscopy in the presence of a magnetic field in atomic samples of rubidium \cite{ciampini_2017,george_2017} and cesium \cite{Staerkind_2022}. These experiments led to metrological determination of $g$-factors with a relative uncertainty reaching $10^{-9}$ level. Being about three orders of magnitude worse, such an error, however, can be significantly reduced in the future for alkali metals, which opens up opportunities for magnetometry and the study of atomic physics in strong magnetic fields \cite{ciampini_2017,george_2017,battesti_2018}.

Based on the analysis given in Ref.~\cite{gfactor2022}, the experiments \cite{ciampini_2017,george_2017,Staerkind_2022} and discussions above, it is of practical interest to study thermal corrections to the $g$-factors of bound electrons in atomic systems with one valence electron, which can lead to results of fundamental importance. A brief theoretical description of the corresponding thermal correction, involving approximations and methods used for numerical calculations, is presented in the next section, whereas discussion and conclusions are given in the last part of our work.

\section{Thermal correction to the bound electron $g$-factor}

Within the framework of the QED approach, one can take into account finite temperature effects using electron and photon Green's functions (propagators) \cite{Dol,DHR}. Additional terms that arise when using the second quantization rules within field theory correspond to the influence of the thermal environment surrounding the considered system. Under laboratory conditions, the electron thermal part is exponentially suppressed, which allows one to consider only the effects induced by the blackbody radiation field. In turn, the presence of two contributions (zero and heated vacuum) in the photon Green's function leads to the possibility of a separate description of the effects at a finite temperature, using the replacement of the propagator defined at the zero vacuum state with the corresponding thermal part in the Feynman diagrams, see Refs.~\cite{ Dol,DHR} and the formalism developed in Refs.~\cite{SLP-QED,S-2020}. 

\begin{figure}%[hbtp]
\centering
\includegraphics[scale=2]{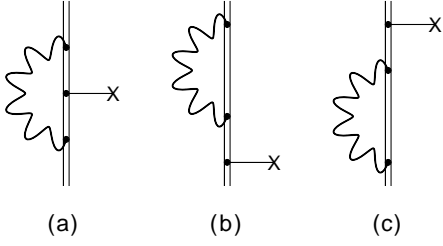}
\caption{Feynman diagrams describing the TQED contributions of the order of $\alpha$ to the bound-electron $g$-factor. The tiny line with the cross indicates interaction with an external magnetic field. The double solid line denotes the bound electron in the Furry picture. The bold wavy line represents the thermal photon propagator, which replaces the ordinary one.}
\label{fig1}
\end{figure}

Details of the QED derivation of the one-loop thermal correction to the bound electron $g$-factor can be found in Ref.~\cite{gfactor2022}. The corresponding Feynman diagrams are shown in Fig.~\ref{fig1}.

Omitting for brevity the analytical derivations made in Ref.~\cite{gfactor2022}, the resulting correction to the energy shift in the dipole approximation can be expressed as (in relativistic units $\hbar=c=1$)
\begin{eqnarray}
\label{1}
\Delta E_{a}^{\mathrm{ver}} =-\frac{2B\mu_{B}}{3\pi}\mathrm{Re}\sum\limits_{\pm}\int\limits_{0}^{\infty}
d\omega \omega^3 n_{\beta}(\omega)\times\qquad
\\
\nonumber
\left[
\sum\limits_{n,n'}
\frac{\langle a| \bm{d} |n\rangle \langle n|v_{z}|n' \rangle \langle n'|\bm{d}|a\rangle}{(E_{a}\pm\omega-E_{n}(1-\mathrm{i}0))(E_{a}\pm\omega-E_{n'}(1-\mathrm{i}0))}
\right],
\end{eqnarray}
\begin{eqnarray}
\label{2}
\Delta E_{a}^{\mathrm{wf}} =-\frac{4B\mu_{B}}{3\pi}\mathrm{Re}\sum\limits_{\pm}\int\limits_{0}^{\infty}d\omega\omega^3 n_{\beta}(\omega) \times \qquad
\\
\nonumber
\left[
\sum\limits_{\substack{n,n'\\n'\neq a}}
\frac{\langle a|\bm{d}|n\rangle \langle n|\bm{d}|n'\rangle \langle n'| v_{z}|a \rangle 
}{(E_{a}\pm\omega-E_{n}(1-\mathrm{i}0))(E_{a}-E_{n'})}
\right.
\\
\nonumber
\left.
- \frac{1}{2} \sum\limits_{n} \frac{\langle a|\bm{d}|n\rangle \langle n|\bm{d}|a\rangle \langle a| v_{z}|a \rangle}{(E_{a}\pm\omega-E_{n}(1-\mathrm{i}0))^2}
\right].
\end{eqnarray} 
The expression (\ref{1}) corresponds to  the vertex part correction given by Feynman graph Fig.~\ref{fig1} (a). The contribution of the wave function correction, Eq. (\ref{2}), is represented by the sum of diagrams (b) and (c) in Fig.~\ref{fig1}. Remaining within the framework of dipole and one-electron approximation, the expressions obtained in this work can be easily generalized to the case of many-electron atomic systems with one valence electron.

In Eqs. (\ref{1}), (\ref{2}) $B$ represents the magnetic field strength, and $\mu_{B}={|e|}/{(2m_{e})}$ ($e$ is the electron charge and $m_e$ is the electron mass) is the Bohr magneton, $n_\beta(\omega)$ is the Planck distribution function, $n_\beta(\omega) = \left(\exp(\omega/k_{\rm B }T)-1 \right)^{-1}$ ($k_{\rm B}$ is the Boltzmann constant, $T$ is the temperature in Kelvin), $\omega$ - thermal photon frequency, $E_n $ is the one-electron energy of the $n$-th state, and $\bm{d}$ is the electric dipole operator $|e|\bm{r}$, ($\bm{r}$ is the radius vector of valence electron). The operator $v_z$ in the nonrelativistic limit can be written as $v_{z} = l_z+2s_z = j_z+s_z$, where $s_z$, $l_z$ and $j_z$ are the projections of the spin, orbital and total momentum onto the $z$ axis, respectively. The sum $\sum\limits_{\pm}$ in Eqs. (\ref{1}), (\ref{2}) denotes the sum of two contributions with the signs "$+$" and "$-$" in the energy denominators of Eqs. (\ref{1}), (\ref{2}). For the atoms with one valence electron summations over quantum numbers $n$ and $n'$ in Eq. (\ref{1}) run over unoccupied one-electron states including positive defined continuum. 

Then, following \cite{gfactor2022}, the $\Delta g_{a}$ correction to the $g$-factor for an arbitrary bound electron state can be obtained from the corresponding energy shift, $\Delta E_{a}$, by the relation
\begin{eqnarray}
\label{3}
\Delta g_{a}=\frac{\Delta E_{a}}{m_{j_a}\mu_{B}B},
\end{eqnarray}
where $m_{j_a}$ is the projection of the total angular momentum $j_a$ for the corresponding atomic state $a$.

 %For hydrogen-like ions with nuclear charge $Z$, the parametric estimate of Eqs. (\ref{1}) and (\ref{2}) can be found by taking into account
 %that in relativistic units $r\sim (m_e\alpha Z)^{-1}$ ($\alpha$ is the fine structure constant), $E_a\sim m_e(\alpha Z)^2$ and 
%$\int\limits_{0}^{\infty}d\omega\omega^{k}n_{\beta}(\omega)\sim(k_{B}^{\mathrm{\;r.u.}}T)^{k+1}$. Then the $g$-factor correction (\ref{3}) is %parametrized as follows
%\begin{eqnarray}
%\label{4}
%\Delta g_{a}\sim \frac{(k_{B}T)^4_{\mathrm{r.u.}}}{\alpha^5 m_{e}^4Z^6}.
%\end{eqnarray}

From the computational point of view, the major problem that occurs in Eqs. (\ref{1}), (\ref{2}) is how to evaluate the double sums over excited states $n$ and $n'$ accurately. For this purpose we follow a technique described in the work \cite{johnson_1987}. According to this work, the spectrum occurring in infinite sums is replaced by a discrete pseudospectrum and then evaluated numerically. The pseudospectrum is constructed by solving the one-electron Dirac equation using two different semi-empirical screening potentials (the Tietz potential and the Green potential, see references in \cite{johnson_1987}), and  local Dirac-Fock (LDF) $V^{N-1}$ potential ($N$ is the total number of atomic electrons) built on a grid, see Ref. \cite{ShabaevPNC} for example. Then numerical evaluation of expressions (\ref{1})-(\ref{2}) was performed by employing the dual-kinetic balance finite basis set method \cite{DKB} with basis functions constructed from B-splines.

The model Tietz potential used in the present calculations (see Ref.~\cite{johnson_1987} and references therein), is given as
\begin{eqnarray}
\label{5}
V^{\rm TP}(r) = -\frac{\alpha}{r}\left[ 1+ \frac{(Z-1)}{(1+t\,r)^2}e^{-\gamma r} \right],
\end{eqnarray}
where $\alpha$ is the fine structure constant,
$t=1.9530\, a_0^{-1}$, $\gamma =0.2700\, a_0^{-1}$
 ($a_0$ is the Bohr radius) for the rubidium ($Z=37$) atom, and $t=2.0453\, a_0^{-1}$, $\gamma =0.2445\, a_0^{-1}$ for the cesium ($Z=55$) atom. The Green Potential is parameterized as follows
\begin{eqnarray}
\label{6}
V^{\rm GP}(r) =  -\frac{\alpha}{r}\left[ 1+ \frac{(Z-1)}{H(e^{r/d}-1)+1} \right],
\end{eqnarray}
with $H=3.48114$, $ 4.46910$ and $d=0.78551\, a_0$, $ 0.89665\, a_0$ for the rubidium and cesium atoms, respectively. 

Before proceeding to evaluate the expressions (\ref{1}) and (\ref{2}), we verify the method described above for calculating energies, dipole matrix elements of allowed transitions, scalar polarizabilities and thermal Stark shifts. First, the Dirac eigenvalues for several low-lying states in the Rb and Cs atoms were calculated with different potentials. The numerical results for binding energies are listed in Table~\ref{tab:1}.
\begin{table}%[hbtp]
\centering
\caption{Dirac eigenvalues in atomic units for  Rb ($Z = 37$) and Cs ($Z = 55$), computed using the Tietz, Green, and local Dirac-Fock (LDF) screening potentials. The last but one column gives the root-mean-square (rms) values obtained for the indicated potentials (the value of the calculated standard deviation, $\sigma$, is indicated in brackets), and the last column shows the values borrowed from Refs.~\cite{WS87,Eriksson_1970,Fendel:07}.}
\begin{tabular}{c c c c c c}
\hline
\hline
State & Tietz & Green & LDF & rms($\sigma$) & Refs. \cite{WS87,Eriksson_1970,Fendel:07}\\
\hline
\hline
& & & Rb &\\
\hline
 $5s_{1/2} $  & -0.15414  & -0.15348 & -0.14113  &-0.1497(73) &-0.15351\\
 $5p_{1/2} $  & -0.09557  & -0.09615 & -0.09124  &-0.0943(27) &-0.09619\\
 $5p_{3/2} $  & -0.09398  & -0.09480 & -0.09027  &-0.0930(24) &-0.09511\\
 $6s_{1/2} $  & -0.06140  & -0.06215 & -0.05927  &-0.0609(15) &-0.06178\\
 $7s_{1/2} $  & -0.03345  & -0.03382 & -0.03269  &-0.0333(6) &-0.03362\\
\hline
\hline
& & & Cs & \\
\hline
 $6s_{1/2} $  & -0.14343 & -0.14310 & -0.13080 &-0.1392(72) & -0.14310\\
 $6p_{1/2} $  & -0.09247 & -0.09223 & -0.08696 &-0.0906(69) & -0.09218\\
 $6p_{3/2} $  & -0.08892 & -0.08915 & -0.08479 &-0.0876(24) & -0.08964\\
 $7s_{1/2} $  & -0.05827 & -0.05901 & -0.05620 &-0.0578(15) & -0.05864\\
 $8s_{1/2} $  & -0.03210 & -0.03248 & -0.03135 &-0.0320(6) & -0.03230\\
\hline
\hline
\end{tabular}
\label{tab:1}
\end{table}

The results in Table~\ref{tab:1} are compiled as follows. The first column indicates specific states, then the second and third columns show the energy values for the Tietz and Green model potentials, defined by the expressions (\ref{5}) and (\ref{6}), respectively, and the results for the local Dirac-Fock potential are listed in the fourth column. The obtained results should be compared with the experimental data \cite{WS87,Eriksson_1970, Fendel:07} given in the fifth column. The values are picked up for the Rb and Cs atoms. 

As it follows from Table~\ref{tab:1}, the binding energies are reproduced by the Tietz and Green potentials with high accuracy (the relative deviation is about one percent, see the $5p_{3/ 2}$ state for Rb and $6p_{3/2}$ state for Cs, or less). In turn, the local Dirac-Fock potential gives energies with a slightly worse accuracy, which is about a few percent for both atoms. Nevertheless, the achieved accuracy of the results for binding energies is more than sufficient for our purposes.

Next, we calculated the reduced matrix elements for the dipole operator of allowed transitions, the values of which are given in Table~\ref{tab:2}.
\begin{table}%[hbtp]
\centering
\caption{Reduced matrix elements in atomic units for allowed transitions calculated
using the Tietz, Green and local Dirac-Fock potentials. The table is organized similarly to Table~\ref{tab:1}.}
\begin{tabular}{c c c c c c}
\hline
\hline
Matrix element & Tietz & Green & LDF & rms($\sigma$) & Refs. \cite{O:81, Antypas,Herold_2012}\\
\hline
\hline
& & &  Rb & \\
\hline
 $\langle 5s_{1/2}||d ||5p_{1/2} \rangle$  &  4.47  & 4.39 & 4.74 & 4.54(18) & 4.231(3)\\
 $\langle 5s_{1/2}||d ||5p_{3/2} \rangle$  &  6.28  & 6.19 & 6.68 & 6.37(26) & 5.978(5)\\
 $\langle 5s_{1/2}||d||6p_{1/2} \rangle$  &  0.45  & 0.40 & 0.39 & 0.41(3) & 0.3236(9)\\
 $\langle 5s_{1/2}||d ||6p_{3/2} \rangle$  &  0.72  & 0.65 & 0.63 & 0.67(5) & 0.5230(8)\\
\hline
\hline
& & & Cs & \\
\hline
  $\langle 6s_{1/2}||d ||6p_{1/2} \rangle$  & 4.85  & 4.74 & 5.12 & 4.91(19) & 4.52(1)\\
 $\langle 6s_{1/2}|d ||6p_{3/2} \rangle$  & 6.77  & 6.64 & 7.20 & 6.87(29) & 6.36(1)\\
 $\langle 6s_{1/2}|d||7p_{1/2} \rangle$  & 0.39  & 0.39 & 0.38 & 0.39(1) & 0.2789(16)\\
 $\langle 6s_{1/2}||d ||7p_{3/2} \rangle$  & 0.79  & 0.76 & 0.72 & 0.76(4) & 0.5780(7)\\
\hline
\hline
\end{tabular}
\label{tab:2}
\end{table}
The numerical results for transitions to the ground states collected in Table~\ref{tab:2} are listed in the same order as in the previous one. The relative deviation for the reduced dipole matrix elements, obtained by comparing our calculations with the experimental data given in the last column of Table~\ref{tab:2}, is worse for higher states $6p_{3/2}$, $7p_{3/2} $ in the rubidium and cesium atoms, respectively, and reaches $25\%$. However, we suppose that this accuracy is sufficient for our further estimations.

Test calculations were carried out also for the scalar polarizabilities $\alpha_{a}^{(0)}$ and the BBR-induced the Stark shifts for a specific state $a$ expressed as
\begin{eqnarray}
\alpha_{a}^{(0)}=\frac{2}{3(2j_{a}+1)}\sum\limits_{n}\frac{|\langle a |\bm{d}|n\rangle|^2}{E_{n}-E_{a}}
,
\end{eqnarray}
\begin{eqnarray}
\Delta E^{\mathrm{BBR}}_{a}
=\frac{2}{3\pi(2j_{a}+1)}\sum\limits_{n}\mathrm{P.V.}\int_{0}^{\infty}d\omega n_{\beta}(\omega)\omega^3
\\\nonumber
\times
|\langle a |\bm{d}|n\rangle|^2
\left[ 
\frac{1}{E_{n}-E_{a}+\omega}
+
\frac{1}{E_{n}-E_{a}-\omega}
\right],
\end{eqnarray}
respcetively.
Evaluation of the properties based on summation over electronic spectra like in Eqs. (\ref{1}), (\ref{2}), thereby allows ones to test the convergence of the generated set of pseudostates. The corresponding values of $\alpha_{a}^{(0)}$ and $\Delta E^{\mathrm{BBR}}_{a}$ are listed in Table \ref{tab:3} and collected as follows.
\begin{widetext}
%\centering
\begin{center}
\begin{table}%[hbtp]
\caption{Polarizabilities $\alpha_a^{(0)}$ in atomic units and BBR-induced Stark shifts $\Delta E^{\mathrm{BBR}}_{a}$ in Hertz at $T=300$ K for the $ns_{1/2}$ states of Rb and Cs atoms. The first line in the 2nd and 4th columns corresponds to calculations performed with the Tietz potential, while the second and third lines to the Green and local Dirac-Fock potentials, respectively. Each picked out line contains the root-mean-square value and the corresponding standard deviation in brackets for the quantities obtained using model potentials.}
\begin{tabular}{c c c c c}
\hline
\hline
State \;\;\;\;\;\;\;\;\;& $\alpha_a^{(0)}$ (this work)\;\;\;\;\;\;\;\;\;& $\alpha_a^{(0)}$,  Refs. \cite{Amini,Weaver}\;\;\;\;\;\;\;\;\;& $\Delta E^{\mathrm{BBR}}_{a}$ (this work)\;\;\;\;\;\;\;\;\;& $\Delta E^{\mathrm{BBR}}_{a}$, Ref. \cite{Farley}\\
 \hline
\hline
& & Rb & & \\
\hline
 $5s_{1/2} $  & 334.91  & 320.1(6) &  -3.85   &  -2.789   \\
              & 332.39  & &  -3.83   &           \\
              & 444.76  & &  -5.13   &            \\ 
              \hline
rms($\sigma$) & 374(64)        & &  -4.31(74)       &     \\
              \hline
 $6s_{1/2} $  & 5153.75  & 5167(22)	 & -63.4  &   -48.17  \\
              & 5034.79  &  & -61.9  &           \\
              & 6033.75  & &  -75.7        &            \\ 
              \hline
rms($\sigma$) & 5426(546)        & &  -67.3(7.6)       &     \\
              \hline
 $7s_{1/2} $  & 31679.8  & 32620(150)	 & -589.4 & -411.7 \\
              & 31671.5  &  & -530.3 &       \\
              & 36650.2  & &  -607.9        &            \\ 
              \hline
rms($\sigma$) & 33416(2872)        & &  -577(41)       &     \\

\hline
\hline
& & Cs & & \\
\hline
 $6s_{1/2} $  & 437.7 & 401.0(6) & -5.02  & -3.589    \\
              & 423.4 &          & -4.86  &           \\
              & 577.9 &          & -6.66      & \\
              \hline
rms($\sigma$) & 485(85)        & &  -5.57(99)       &     \\
              \hline
 $7s_{1/2} $  & 6400.3 & 6111(23) & -79.6 & -59.67    \\
              & 6073.6 &          & -75.3 &           \\
              & 7302.2 &          & -92.7      &\\              
              \hline
rms($\sigma$) & 6612(636)        & &  -82.9(9.1)       &     \\
              \hline
 $8s_{1/2} $  & 35853.6 & 38370(380) & -589.4 & -477.4\\
              & 34740.9 &            & -573.3 &       \\
              & 39135.8 &            & -641.2      &\\        
              \hline
rms($\sigma$) & 36624(2285)        & &  -602.0(35.5)       &     \\
\hline
\hline
\end{tabular}
\label{tab:3}
\end{table}
\end{center}
\end{widetext}
The first column specifies the considered states, then the second column indicates the values obtained with the Tietz, Green and LDF potentials in sublines, respectively. The third column shows the values borrowed from Refs.~\cite{Amini,Weaver} for comparison with our results. The same for the Stark shifts caused by BBR, the values of which are given in the fourth and fifth columns. The Tietz and Green potentials show an accuracy of $10\%$ for the scalar polarizabilities in cesium and somewhat less in rubidium compared to the results of Refs.~\cite{Amini,Weaver}. In turn, the LDF model potential gives results of about $15\%$ in cesium and $30\%$ in rubidium for lower states, getting better for higher states.

Summarizing the obtained results demonstrate satisfactory agreement between the theoretical binding energies in the Rb and Cs atoms and the experimental ones. In turn, the transition matrix elements estimated using different potentials deviate in the second digit from the values determined experimentally with an accuracy of $10\%$ \cite{Antypas}. In fact, the detected deviations of the matrix elements can be explained by the accuracy of our calculations with model potentials restricted by the zero order contribution of many-body perturbation theory (MBPT). Inclusion of the next order corrections within the MBPT approach greatly improves the accuracy \cite{johnson_1987}, but complicates the calculations. However, since we would like to carry out only estimates of the thermal corrections to $g$-factor, expressed by the equations (\ref{1}), (\ref{2}) and (\ref{3}), then the obtained values are sufficient for our purposes. Finally, according to Table~\ref{tab:3}, we determine the accuracy of our calculations as the worst deviation between $\Delta E^{\mathrm{BBR}}_{a}$ calculated here and in Ref.~\cite{Farley}, i.e. about of $30\%$ (see also Table~\ref{tab:4}). 

%The numerical results for the thermal correction to the $g$-factor in cesium and rubidium atoms, corresponding to the Feynman diagrams depicted in Fig.~\ref{fig1}, are given in Table~\ref{tab:4} for the Tietz, Green and LDF model potentials (in sublines, respectively).

While having an acceptable precision for our purposes, present calculations imply the correct order of magnitude, which can be found for the expressions (\ref{1}) and (\ref{2}). 
%The BBR field is defined with the relation \cite{Itano}:
%\begin{eqnarray}
%\label{7}
%\langle E^2(\omega)\rangle=\langle B^2(\omega)\rangle=\frac{8\alpha^3}{\pi}\frac{\omega^3}{e^{\frac{\omega}{k T}}-1},
%\end{eqnarray}
%where squared average of the electric field $\langle E^2(\omega)\rangle$ or the magnetic field $\langle %B^2(\omega)\rangle$ is expressed through the Planck distribution law, $n_\beta(\omega)$.
The numerical results for the one-loop thermal self-energy correction to the bound-electron $g$-factor, see Fig.~\ref{fig1}, given by the formula (\ref{3}) are collected in Table~\ref{tab:4} for several low-lying states of the Rb and Cs atoms.
 \begin{table}%[hbtp]
\caption{Thermal one-loop corrections to the bound electron $g$-factor for low-lying states of the Rb and Cs atoms, calculated using the Tietz (first line), Green (second line) and local Dirac-Fock (third line) potentials at different temperatures of the BBR background. Each picked out line contains the root-mean-square value and the corresponding standard deviation in brackets for the thermal correction $\Delta g$ determined by Eq. (\ref{3}).}
\begin{center}
%\centering
%{\renewcommand{\arraystretch}{1.2}
\setlength{\extrarowheight}{2pt}
\begin{tabular}{c c c}

\hline
\hline
 $\Delta g$ & 300 K & 1000 K\\
\hline
& Rb & \\
\hline
 $\Delta g[5s_{1/2}] $  &  0.91$\times 10^{-14}$ &  1.34$\times 10^{-12}$ \\
            &  0.92$\times 10^{-14}$ &  1.37$\times 10^{-12}$\\
            &  1.42$\times 10^{-14}$ &  2.30$\times 10^{-12}$\\
            \hline
rms($\sigma$) & 1.11(29)$\times 10^{-14}$  & 1.73(55)$\times 10^{-12}$\\
            \hline
 $\Delta g[5p_{1/2}] $  &  1.29$\times 10^{-14}$ &  2.55$\times 10^{-12}$\\
             &  1.25$\times 10^{-14}$ &  2.48$\times 10^{-12}$\\
             &  1.66$\times 10^{-14}$ &  3.35$\times 10^{-12}$\\
            \hline
rms($\sigma$) & 1.41(23)$\times 10^{-14}$  & 2.82(48)$\times 10^{-12}$\\
            \hline
 $\Delta g[5p_{3/2}] $  &  7.31$\times 10^{-15}$ &  1.49$\times 10^{-12}$\\
             &  6.98$\times 10^{-15}$ &  1.41$\times 10^{-12}$\\
             &  8.95$\times 10^{-15}$ &  1.82$\times 10^{-12}$\\
            \hline
rms($\sigma$) & 7.8(1.1)$\times 10^{-15}$  & 1.58(22)$\times 10^{-12}$\\
% $\Delta g[6s_{1/2}] $  &  1.12$\times 10^{-13}$ &  1.44$\times 10^{-12}$\\
%             &  6.68$\times 10^{-13}$ &  -2.59$\times 10^{-11}$\\
%             &  5.61$\times 10^{-14}$ &   2.86$\times 10^{-12}$\\
% $\Delta g[7s_{1/2}] $  &  1.24$\times 10^{-11}$ &  -7.59$\times 10^{-10}$ \\ 
%                        &   1.39 $\times 10^{-11}$      &    -9.17$\times 10^{-10}$ \\
%                        &   1.22 $\times 10^{-11}$      &    -7.77$\times 10^{-10}$ \\
\hline

\hline
\hline
& Cs & \\
\hline
 $\Delta g[6s_{1/2}] $  & 8.59$\times 10^{-16}$ &  1.67$\times 10^{-13}$\\
                        & 7.12$\times 10^{-16}$ &  1.40$\times 10^{-13}$\\
                        & 9.22$\times 10^{-16}$ &  2.17$\times 10^{-13}$\\
            \hline
rms($\sigma$) & 8.4(1.1)$\times 10^{-16}$  & 1.78(39)$\times 10^{-13}$\\
            \hline
 $\Delta g[6p_{1/2}] $  & 1.47$\times 10^{-14}$ &  2.88$\times 10^{-12}$\\
                        & 1.50$\times 10^{-14}$ &  2.98$\times 10^{-12}$\\
                        & 2.11$\times 10^{-14}$ &  4.31$\times 10^{-12}$\\
            \hline
rms($\sigma$) & 1.72(36)$\times 10^{-14}$  & 3.45(80)$\times 10^{-12}$\\
            \hline
 $\Delta g[6p_{3/2}] $  & 9.57$\times 10^{-15}$ &  1.96$\times 10^{-12}$\\
                        & 9.57$\times 10^{-15}$ &  1.96$\times 10^{-12}$\\
                        & 1.24$\times 10^{-14}$ &  2.52$\times 10^{-12}$\\
                        
            \hline
rms($\sigma$) & 1.06(16)$\times 10^{-14}$  & 2.16(32)$\times 10^{-12}$\\
% $\Delta g[7s_{1/2}] $  & 1.08$\times 10^{-12}$ & -9.03$\times 10^{-11}$\\
%                        & 0.99$\times 10^{-12}$ & -7.76$\times 10^{-11}$\\
%                        & 1.34$\times 10^{-12}$ & -1.28$\times 10^{-10}$\\
% $\Delta g[8s_{1/2}] $  & 1.28$\times 10^{-12}$ & -1.68$\times 10^{-10}$\\
%                        & 1.36$\times 10^{-12}$ & -1.64$\times 10^{-10}$\\ 
%                        & 1.39$\times 10^{-11}$ & -1.01$\times 10^{-9}$\\
\hline
\hline
\end{tabular}
%}
\end{center}
\label{tab:4}
\end{table}
The results for the temperatures $T=300$ and $T=1000$ K are given to illustrate the behavior of the correction (\ref{3}) as $T^4$.

\section{Discussion and conclusions}

%The precise determination of the $g$-factor is of fundamental importance in physics, serving to verify calculations of atomic structure, to determine fundamental physical constants, such as, e.g., the fine structure constant or the mass of an electron and etc \cite{werth:18:aamop,nature2014}. The experiments carried out to determine the $g$-factor of a bound electron in hydrogen-like ions have reached an extremely high precision in recent years, providing the basis for theoretical calculations of various effects at the level of experimental accuracy and below. Latterly, the correction to the $g$-factors induced by the blackbody radiation field was discussed for light hydrogen-like ions in \cite{gfactor2022}. It has been found that thermal induced corrections to the $g$-factors of excited states may play a significant role in the near future reaching the level of $10^{-12}$ at room temperature.

%Thermal corrections to the energies of atomic states or to physical quantities determined with high accuracy are of particular interest for heavy neutral atoms. First of all, this is dictated by the 
The construction and corresponding development of atomic clocks operating on alkali metals, such as, for example, Rb and Cs, made it possible to carry out spectroscopic measurements of transition frequencies with accuracy leading to date. Achieving extraordinary precision, atomic clocks serve a fiducial point for measuring other physical quantities. In particular, one of the many applications of precision optical spectroscopy was recently demonstrated in Refs.~\cite{ciampini_2017,george_2017,Staerkind_2022}, where the atomic response to magnetic fields was measured. Experiments \cite{ciampini_2017,george_2017,Staerkind_2022} showed the possibility of determining the $g$-factor with a metrological accuracy of $10^{-9}$, as well as ways to improve it further. 

One of the factors limiting the experimental precision of atomic clocks is the effects induced by the BBR, which requires accurate theoretical calculations of the BBR-Stark shifts and experimental control of the thermal environment \cite{Nicholson}. The latter leads to the conclusion that any physical quantity observed with such a degree of accuracy is needed to be considered with the account for thermal effects. Based on the theoretical description given in Ref.~\cite{gfactor2022}, we have evaluated the thermal correction shown in Fig.~\ref{fig1} to the $g$-factor.

The semi-empirical and local Dirac-Fock potentials were used as the simplest ones for calculating Eqs. (\ref{1}), (\ref{2}). To establish the accuracy of our results, different potentials were employed to estimate the zero-order binding energies in Rb and Cs, see Table~\ref{tab:1}. Then the dipole matrix elements were calculated, Table~\ref{tab:2}. The combination of these quantities makes it possible to estimate the polarizabilities and the corresponding BBR-Stark shifts, the numerical results of which are given in Table~\ref{tab:3}. It can be seen from these tables that the calculations agree satisfactorily with the experimental data. We set a rough estimate of the accuracy of our calculations as $30\%$, which gives at least the correct order of magnitude.

Finally, the correction to the $g$-factor of the particular states in rubidium and cesium atoms was obtained from Eq. (\ref{3}) using the expressions (\ref{1}), (\ref{2}), see Table~\ref{tab:4}. Despite the fact that the values of this correction are still less than the experimental error achieved in determining the $g$-factor, it is important to compare it with other effects that contribute to the uncertainty of the $g$-factor and are widely discussed in the literature. In particular, the calculated thermal corrections even at room temperature exceed or are of the same order of magnitude with the QED corrections due to fluctuations of the muonic vacuum. The latter recently were calculated in Ref.~\cite{belov2016muonic}. In addition, the current hadronic and QED uncertainties in theoretical value of the $g$-factor are at the level of $10^{-14}$ \cite{Richard,Morel,Parker}, which roughly corresponds to the deviation found in the present study. Recently in Ref.~\cite{Cohen} it was shown that the gravitational corrections to the electron $g$-factor also reach this magnitude. Finally, an indirect determination of the $g$-factor in the Rb atom was carried out by determining the value of the fine structure constant and, as a consequence, the anomalous magnetic moment of the electron, which gives the difference between theoretical and experimental values at the level of $(40\pm 89)\times 10^{- 14}$ \cite{Bouchendira}. The later is of an order of magnitude larger than the found thermal corrections, see Table~\ref{tab:4}.

In conclusion, we note that the value of thermal correction $\Delta g$ is orders of magnitude less than the current accuracy of indirect determination of $g$-factor in the Cs atom via measured value of the fine structure constant \cite{Staerkind_2022}. Considering the relative uncertainty of the order of $10^{-17}$ in determining the transition frequencies in the Rb and Cs atoms, a significant improvement of the $g$-factor determination, which can compete with hydrogen-like systems, can be expected in the near future.

%{\it Acknowledgements.}
%\env{acknowledgments}
\acknowledgments This work was supported by Russian Foundation for Basic Research (grant 20-02-00111). The authors are grateful to prof. I. I. Tupitsyn for valuable discussion.

\bibliographystyle{apsrev4-1}
\bibliography{mybibfile}% Produces the bibliography via BibTeX.

\end{document}